
%
%
%
\documentstyle{amsppt}
\newread\epsffilein    
\newif\ifepsffileok    
\newif\ifepsfbbfound   
\newif\ifepsfverbose   
\newdimen\epsfxsize    
\newdimen\epsfysize    
\newdimen\epsftsize    
\newdimen\epsfrsize    
\newdimen\epsftmp      
\newdimen\pspoints     
\pspoints=1bp          
\epsfxsize=0pt         
\epsfysize=0pt         
\def\epsfbox#1{\global\def\epsfllx{72}\global\def\epsflly{72}%
   \global\def\epsfurx{540}\global\def\epsfury{720}%
   \def\lbracket{[}\def\testit{#1}\ifx\testit\lbracket
   \let\next=\epsfgetlitbb\else\let\next=\epsfnormal\fi\next{#1}}%
\def\epsfgetlitbb#1#2 #3 #4 #5]#6{\epsfgrab #2 #3 #4 #5 .\\%
   \epsfsetgraph{#6}}%
\def\epsfnormal#1{\epsfgetbb{#1}\epsfsetgraph{#1}}%
\def\epsfgetbb#1{%
%
%
\openin\epsffilein=#1
\ifeof\epsffilein\errmessage{I couldn't open #1, will ignore it}\else
%
%
   {\epsffileoktrue \chardef\other=12
    \def\do##1{\catcode`##1=\other}\dospecials \catcode`\ =10
    \loop
       \read\epsffilein to \epsffileline
       \ifeof\epsffilein\epsffileokfalse\else
%
%
          \expandafter\epsfaux\epsffileline:. \\%
       \fi
   \ifepsffileok\repeat
   \ifepsfbbfound\else
    \ifepsfverbose\message{No bounding box comment in #1; using defaults}\fi\fi
   }\closein\epsffilein\fi}%
%
%
\def\epsfclipstring{}
\def\epsfsetgraph#1{%
   \epsfrsize=\epsfury\pspoints
   \advance\epsfrsize by-\epsflly\pspoints
   \epsftsize=\epsfurx\pspoints
   \advance\epsftsize by-\epsfllx\pspoints
%
%
   \epsfxsize\epsfsize\epsftsize\epsfrsize
   \ifnum\epsfxsize=0 \ifnum\epsfysize=0
      \epsfxsize=\epsftsize \epsfysize=\epsfrsize
      \epsfrsize=0pt
%
%
     \else\epsftmp=\epsftsize \divide\epsftmp\epsfrsize
       \epsfxsize=\epsfysize \multiply\epsfxsize\epsftmp
       \multiply\epsftmp\epsfrsize \advance\epsftsize-\epsftmp
       \epsftmp=\epsfysize
       \loop \advance\epsftsize\epsftsize \divide\epsftmp 2
       \ifnum\epsftmp>0
          \ifnum\epsftsize<\epsfrsize\else
             \advance\epsftsize-\epsfrsize \advance\epsfxsize\epsftmp \fi
       \repeat
       \epsfrsize=0pt
     \fi
   \else \ifnum\epsfysize=0
     \epsftmp=\epsfrsize \divide\epsftmp\epsftsize
     \epsfysize=\epsfxsize \multiply\epsfysize\epsftmp
     \multiply\epsftmp\epsftsize \advance\epsfrsize-\epsftmp
     \epsftmp=\epsfxsize
     \loop \advance\epsfrsize\epsfrsize \divide\epsftmp 2
     \ifnum\epsftmp>0
        \ifnum\epsfrsize<\epsftsize\else
           \advance\epsfrsize-\epsftsize \advance\epsfysize\epsftmp \fi
     \repeat
     \epsfrsize=0pt
    \else
     \epsfrsize=\epsfysize
    \fi
   \fi
%
%
   \ifepsfverbose\message{#1: width=\the\epsfxsize, height=\the\epsfysize}\fi
   \epsftmp=10\epsfxsize \divide\epsftmp\pspoints
   \vbox to\epsfysize{\vfil\hbox to\epsfxsize{%
      \ifnum\epsfrsize=0\relax
        \includegraphics{#1}%
      \else
        \epsfrsize=10\epsfysize \divide\epsfrsize\pspoints
        \includegraphics{#1}%
      \fi
      \hfil}}%
\global\epsfxsize=0pt\global\epsfysize=0pt}%
%
%
{\catcode`\%=12 \global\let\epsfpercent=
%
%
\long\def\epsfaux#1#2:#3\\{\ifx#1\epsfpercent
   \def\testit{#2}\ifx\testit\epsfbblit
      \epsfgrab #3 . . . \\%
      \epsffileokfalse
      \global\epsfbbfoundtrue
   \fi\else\ifx#1\par\else\epsffileokfalse\fi\fi}%
%
%
\def\epsfempty{}%
\def\epsfgrab #1 #2 #3 #4 #5\\{%
\global\def\epsfllx{#1}\ifx\epsfllx\epsfempty
      \epsfgrab #2 #3 #4 #5 .\\\else
   \global\def\epsflly{#2}%
   \global\def\epsfurx{#3}\global\def\epsfury{#4}\fi}%
%
%
\def\epsfsize#1#2{\epsfxsize}
%
%

\nologo

\hsize 32pc
\vsize 50pc
\emergencystretch=100pt

\magnification =1200
\pageheight{ 7.5 in}
\baselineskip=25pt plus 2pt

%
%

\def\ms{{\medskip}}
\def\del{{\partial}}
\def\k{{\kappa}}
\def\t{{\tau}}

\def\slr{{\Cal R}}
\def\sll{{\Cal L}}
\def\slh{{\Cal H}}

\def\sls{{\Cal S}}

\def\slp{{\Cal P}}

\def\slz{{\Cal Z}}

\def\lb{{\{}}
\def\rb{{\}}}

\topmatter

\title
The planar filament equation
\endtitle
\author
 Joel Langer  and  Ron Perline
\endauthor
\affil
Dept. of Mathematics, Case Western Reserve University \\
Dept. of Mathematics and Computer Science, Drexel University
\endaffil
\abstract
{The planar filament equation and its relation
to the modified Korteweg-deVries equation are
studied in the context of Poisson geometry.  The
structure of the planar filament equation is
shown to be similar to that of the 3-D localized
induction equation,
previously studied by the authors.}
\endabstract
\endtopmatter

{\bf Remarks:  The authors would appreciate it if anyone who downloads
this file from the nonlinear science preprint archives would send an
acknowledgement to: rperline\@mcs.drexel.edu.  We are interested in tracking
the archive usage.  Thanks!}
\ms
\ms

\subheading{Introduction}
In this paper we study the {\it planar filament equation}
$$ \gamma_t  =
{1 \over 2}{\kappa^2}T + \kappa_s N , \leqno(\hbox{PF})$$
where $\gamma(s,t)$ denotes an evolving planar curve,
parameterized by arclength $s$,
$\kappa$ is its  curvature, $T$ its unit tangent, and $N$ its unit normal.
(PF) is a particular example of a geometric evolution
equation on planar curves.
By this we mean an evolution equation  such
that the velocity of any point on the curve is given in terms of the
curve's geometric invariants.  Such equations have been
studied in a variety of applied contexts, as well as for their
intrinsic geometric interest.  The equation
$ \gamma_t  = V_0 ( 1 - \epsilon \kappa) N$
appears in combustion theory and crystal growth
([Lan],[M1r]).
The {\it curve shortening flow} $ \gamma_t =
\kappa N$ has recently been the focus of intensive  study by
geometers (see, for example  [Ga] and [Ga-H]).
One impetus for its study has been its relation to the
construction of geodesics in Riemannian geometry; but other recent
research  has been concerned with the  analytical properties of the flow
on curves in $R^2$, where of course the study of geodesics is not an
issue.  The curve shortening flow can be thought of as (minus) the
gradient flow of the length functional $\sll (\gamma) = \int_\gamma {\,
ds} $ on the appropriate space of curves.
As we shall see, (PF) can be viewed as a {\it Hamiltonian evolution equation,
with $\sll$ as the Hamiltonian}.
Thus (PF) is a very natural equation to study. (PF)
has already appeared in the mathematical literature in  several
contexts. It appears implicitly in [Ch-T], which is a discussion of
integrable foliations by curves of surfaces of constant curvature.  Also,
in [La 1-2], (PF) is mentioned as one of a list of evolution
equations of curves related to nonlinear evolution equations solved by
the two-component inverse-scattering method.  For a more recent discussion,
see [G-P].  All of these authors point out the close connection between
(PF) and the {\it modified Korteweg-deVries equation}
$$ u_t = u_{sss} + {3 \over 2} u^2 u_s\, , \leqno(\hbox{mKdV})$$
one of the well known examples of a completely integrable equation;
we will later describe the connection in detail.
In this paper, we wish to emphasize the integrability of (PF)
itself and think of
(PF) as  {\it the} geometric realization of (mKdV);  somewhat facetiously,
(PF) is ``mKdV for geometers." (PF) may well have the distinction of
being the simplest geometric evolution equation which is integrable.

Our interest in (PF) came from an observation made in the course
of our recent work on the
{\it localized induction equation}
$$\gamma_t =  \kappa B.
\leqno(\hbox{LIE})$$ %
Here, $\gamma(s,t)$ is  (for fixed $t$) a space curve parameterized
by arclength $s$, with curvature $\kappa$, torsion $\tau$ and Frenet frame
$T,N,B$. (LIE) is a simplified model of vortex filament evolution in fluid
mechanics.
(LIE) was known to be Hamiltonian on the  appropriate space of curves([M-W]).
In [L-P 2], we established that (LIE) is in fact a completely
integrable Hamiltonian equation; in particular, there exists an infinite
sequence
of commuting vectorfields, beginning with (LIE).  We list the first few such
vectorfields:
$$\eqalign{
X_{-2} &=  \k B , \cr
X_{-1} &=  {1 \over 2} \k ^2 T + \k ' N + \k \t B , \cr
X_{0}\ \   &=  \k ^2 \t T + (2 \k ' \t + \k \t ') N \cr
       &+  (\k \t ^2 - \k '' - {1 \over 2} \k ^3 ) B, \cr
X_{1}\  \  &= (-\k \k '' + {1 \over 2} (\k ')^2 + {3 \over 2} \k ^2 \t ^2 -
{3 \over 8} \k
^4)T \cr
       &+ ( - \k ''' + 3 \k \t \t ' + 3 \k ' \t ^2 - {3 \over 2} \k ^2 \k ') N
\cr
       &+  ( \k \t ^3 - 3 (\k ' \t ) ' -  {3 \over 2} \k ^3 \t - \k \t '') B.
\cr
}$$

Here $'$ denotes differentiation with respect to $s$.  We note that (LIE) is
the
first vectorfield appearing in the sequence.  The numbering system is explained
in [L-P 2] and need not concern us here.  What is of interest is that the
{\it odd} vectorfields listed preserve {\it planarity} of curves.  If $\gamma$
is planar, then of course $\tau(s)=0$ for all values of $s$.  But in that case,
the coefficients of $B$ in $X_{-1} \hbox{\ and \ }X_{1}$ vanish by inspection.
The resulting vectorfields only have tangential and normal components, thus
preserving planarity of the initial curve.  Also note that $X_{-1}$ restricted
to planar curves is just (PF).

This line of reasoning suggests that perhaps all the odd vectorfields in our
infinite list are planarity-preserving, and that this infinite sequence of
vectorfields defined on planar curves, starting with (PF) itself, is an
integrable system. In this paper, we illustrate the basic aspects
of the structure of this integrable hierarchy, as well as its relation
to the (mKdV) hierarchy.  It is also interesting to observe that
(PF) ``inherits" its structure from (LIE); upon occasion, we will refer
to [L-P 2]  for the aspects of the structure of the (LIE) hierarchy which
we will find useful.
\medskip

\subheading{Structure of the planar filament equation}
It is helpful
to make precise the space of curves we shall be  working with:
We set $AL = \lb \gamma: (-\infty , \infty) \rightarrow {\Bbb R}^2:$
$ \gamma$
is arc-length-parameterized, and is
asymptotic to the $x$-axis $\rb$.  To say $\gamma$ is asymptotic to the
$x$-axis means that  there exist constants $\lambda_+ , \lambda_-$ such that

$$\lim_{s \to {\pm \infty}} \  [ s e_1 -  \gamma (s) ] = \lambda_{\pm}
e_1 \ ,\leqno(2.1)$$
where $e_1 = (1,0)$.

Within $AL$, there is a distinguished subspace of
 {\it balanced}
asymptotically linear curves $BAL = \{ \gamma \in AL : 0 = \Lambda _+ =
\lambda _+ + \lambda _- \}$.  Elements of $BAL$ correspond to convenient
parameterizations of asymptotically linear curves. This is reflected in
the fact that
there is a simple characterization of the tangent space to $BAL$:
if $\gamma \in BAL$ and $W = fT + gN$ is a vector field along
$\gamma$ ($T,N$ is the Frenet frame along $\gamma$) then
$$T_{\gamma} BAL =  \{W: f = \sls (g \kappa ) \} \ , \leqno(2.3)  $$
where $\sls$ denotes the antidifferentiation operator
$$\sls (f)(s)  =  ({1 \over 2})\  [ \  \ {\int_{ - \infty}^s f(u) {\ du \ }}
- {\int_{s}^\infty f(u) {\ du \ }} \  ]
$$

We now consider the curvature correspondence which assigns to a curve
$\gamma \in BAL$ its curvature function $\kappa(s)$.  Of course, this
correspondence has been considered in many contexts; but it is worth
noting that we can consider this a special case of the {\it Hasimoto
transformation} (introduced in [Has2] and subsequently studied in [L-P 1-2])
which assigns to a  space curve a complex function via the formula
$\gamma \rightarrow \slh(\gamma) = \kappa(s)e^{i \int \tau(u) du}$
 (for planar curves $\tau$ vanishes and we are just left with
$\kappa$).  It
is for this reason that we denote the curvature correspondence for planar
curves by $\slh_P$, and informally refer to it as the {\it planar
Hasimoto transformation}.

Our first task is to compute the differential of $\slh_P$ , considered
as a map from $BAL$  to the space of functions.  To do
so, we recall some  elementary formulas for the variation of geometric
invariants along a curve ([L-P 2]):

\proclaim{Proposition} Denote by $\gamma = \gamma (w,u): (-\epsilon, \epsilon)
\times
(a,b) \rightarrow {\Bbb R}^3$ a one-parameter family of planar curves.
If $W ={{ \del \gamma} \over {\del w }}(0,u)$ is the variation vector
field along  $\gamma$,
 and $\gamma$ has speed $v = |{{ \del \gamma} \over {\del u }} |$,
curvature
$\kappa$, then $v$ and $\kappa$ vary according to:
$$\leqalignno{W(v) &= \ <W' , T> v = -\alpha v, \quad \alpha = - < W', T >
 &(a) \cr
W(\kappa) &= \ <W'' , N> - 2 <\kappa W' , T> &(b) \cr}$$
\endproclaim
Again, $'$ denotes derivative with respect to the arclength parameter along
$\gamma$.

{}From  formula ($a$), it is immediate that if $W = fT + gN$
has coefficients satisfying $f'(s) = g(s) \kappa(s)$, then the variation
of $v$ along $W$ is $0$; we call such vector fields {\it locally arclength
preserving}, for obvious reasons.  Observe that elements of $T_{\gamma} BAL$
satisfy this condition.

Now let $\gamma$ be an element of $BAL$, and $W \in  T_{\gamma} BAL$.
Formula ($b$) gives a simple formula for the variation of $\kappa$ along
$W$:
$$W(\kappa) \ = \  {d\slh_P (W)}  = \ <W'' , N>,$$

{\it thus giving a formula for the differential of the planar Hasimoto
transformation}.

As a first application, let $W = {1 \over 2} \kappa ^2 T + k' N $ which is
obviously in $T_\gamma BAL$.  Then by the Frenet equations, we obtain :
$$\eqalign{
W' &=  (\kappa '' + {1 \over 2} \kappa ^3 ) N  ,\cr
\hbox{\ thus \ }
 {d\slh_P (W)} \  &= \ <W'' , N > \  =  \  \kappa ''' + {3 \over 2}
\kappa ^2 \kappa '
\cr }$$

An immediate consequence of
this last calculation is the relation of (PF) to
(mKdV): if we have a solution $\gamma (s,t)$ to (LIE), then $\kappa (s,t)$
satisfies (mKdV).  Conversely, given a solution $\kappa (s,t)$ to (mKdV),
we can reconstruct the unit tangent vector $T(s,t)$ to $\gamma$ by
computing $\theta = \int^s \kappa ds$, and setting $T = (cos(\theta),
 sin(\theta))$; the curve $\gamma$ itself is obtained by one further
antidifferentiation of $T$.

Soliton solutions for (mKdV) are well-known ([Hir]); these give us
explicit formulas for $\kappa(s,t)$.  At the end of this paper, we
show plots of an evolving 2-soliton curve with curvature $\kappa$
satisfying (mKdV),
as viewed from a moving
frame of reference so that the ``lowest" point of the curve is centered with
respect to the viewing box.

Going back to the formula for the differential of the planar Hasimoto
map, we interpret this formula in geometric terms.
Let us return for a moment to
the localized induction equation.  The   (LIE)  hierarchy
enjoys a number of properties common to
integrable systems. In particular, there is a linear {\it recursion
operator} $\slr$
which generates the commuting vector fields and is given by
the formula
$$
 \slr W = - \slp (T \times W'),  \ \hbox{where} \
 W = fT + gN + hB \ \hbox{and} \
\slp W = \sls (g \kappa) T + gN + hB \ .  $$

By stating that $\slr$ generates the (LIE) hierarchy, we mean more
precisely that $X_k = \slr^{k+2}X_{-2} \, , \,
k=0,1,2,\, \dots$; this is proved in [L-P 2] and
the reader can easily check the first few instances of this formula
by considering the list of $X_i$ already presented from the (LIE) hierarchy.

One can  compute $\slr ^2 W$ to obtain
$$\slr ^2 W =  - \slp (W'' + \sls (<W' , \kappa B>) \kappa B).$$
If we now consider a planar curve $\gamma$ and a planar field $W$ along
$\gamma$, this formula simplifies to
$\slr ^2 W = - \slp (W'')$.  To complete the connection with the differential
for the planar Hasimoto map, we introduce the operator $\slz$ mapping
$$\slz : W  \rightarrow \   <W,N> . $$ Then we observe that
$d\slh_P(W) =-\slz \slr ^2 (W)$.

As just stated, we know that
$\slr$ satisfies $\slr ^2 X_k = X_{k+2}, \ k=-2,-1, \dots \ $.  From the
formula for $\slr ^2$ given above, it is clear that if $\gamma$ is a planar
curve with a planar field $W = fT + gN$ along it, then $\slr ^2$ preserves
the planarity of $W$.  In particular, the field $X_{-1} =
{{\kappa ^2} \over 2}T + \kappa _s N + \kappa \tau B$ restricts to
${{\kappa ^2} \over 2}T + \kappa_s N$ along a planar curve $\gamma$,
and hence $all$ the odd flows in (LIE) hierarchy preserve planarity
(The fact that these planar vectorfields commute follows immediately
from the fact that they are restrictions of commuting vectorfields
defined along space curves).
We call the restriction of $\slr^2$ to planar vectorfields along planar
curves the {\it planar recursion operator} and denote it by $\slr_P$. We
can now express the differential formula for the planar Hasimoto transformation
via $d\slh_P(W)= -\slz \slr_P (W)$.

Now that we have demonstrated that the sequence of vectorfields
$X_{-1}, X_1, X_3, \dots $ induce evolution equations on planar curves,
we can discuss
their hamiltonian nature.
As is well known (see [F-T]) there exist an infinite sequence of
conserved quantities for the (mKdV) hierarchy, defined in terms
of integrals of polynomials in $u$ and its derivatives:
$$\tilde I_0 = \int {1 \over 2} u^2 ds, \ \ \tilde I_2 =
\int ({1 \over 2} (u')^2 - {1 \over 8}u^4 ) ds \, , \dots $$

Given the relation between (PF) and (mKdV) previously demonstrated,
it is a simple consequence of the chain rule that these conserved
quantities ``pull back" to give invariants for (PF) (simply replace
$u$ by $\kappa$).

These invariants for (PF):
$I_0 = \int {1 \over 2} \k^2 ds, \ \ I_2 =
\int ({1 \over 2} (\k')^2 - {1 \over 8}\k^4 ) ds,  \, \dots $
are closely related the invariants $I_i$ of the (LIE) hierarchy
discussed in [L-P 2]: they are the restriction to planar curves
of the {\it even} invariants of (LIE); the odd invariants, restricted
to planar curves, vanish.
 Again invoking [L-P 2], we know that the full (LIE)
hierarchy can be put in hamiltonian form using these
invariants: $X_i = J \nabla I_i$, where
the $I_i$ are the invariants of the (LIE)
hierarchy, $\nabla$ denotes
formal $L^2$ gradient, and
$J$ denotes the skew operator
$J(W) = - \slp (T \times W)$.  But the (LIE) vectorfields
are also hamiltonian with respect to a second operator:
$X_i = K \nabla I_{i-1}$, where $K = J(d/ds)J$.  If $\gamma$ is planar
with planar field $W= fT + gN$, then $K(W) = \sls (\kappa g')T + g'N$.
In particular, $X_{2k+1} = K \nabla I_{2k}, \ k=0,1,2, \dots$.  From
our previous discussion, all the  objects in this last equation make
sense for planar curves: $I_{2k}$ restricts nontrivially to planar curves,
with gradient involving components in the $T$ and $N$ directions; $K$
preserves
planarity;  and the $X_{2k+1}$ are also defined on planar curves.

This demonstrates the hamiltonian nature of all the vectorfields in the (PF)
hierarchy, except (PF) itself! This is because the hamiltonian for (PF)
is {\it not} one of the invariants induced by (mKdV), but rather
the (renormalized) length functional $\sll$ which can be defined on
our space of curves $BAL$ via the formula
$\sll(\gamma) = \lambda_+ -\lambda_-$ (If we had been dealing
with periodic curves, then the hamiltonian would have been the
ordinary length functional).  One can  show that  $\nabla \sll = -\k N$
(as one would expect from the ordinary length functional) and
$K \nabla \sll = -( {1 \over 2}\kappa^2 T  + \k' N)$.  Thus, up to a minus
sign, we obtain (PF).

Finally, we discuss how one can interpret $\slh_P$ as a Poisson map.
To do so, we recall ([F-T], [L-P 2]) that (mKdV) is also hamiltonian;
it can be written as $u_t= \tilde K \nabla F$, where $\tilde K$ is the
skew operator $d \over {ds}$, and $F$ is just (minus) ${\tilde I}_2$.
The (mKdV) hierarchy of commuting evolution equations is generated
by the recursion operator ${\tilde \slr }(u,v) =
v'' + u^2 v' + u' \sls (u v)$ (in this discussion, our notational conventions
are different from those we used in [L-P 2]).
There exists an infinite sequence of Poisson brackets  associated with
$\tilde K$ and $\tilde \slr$:
For any two functionals
$F = \int f(u,u',u'', \, \dots) ds$ and $G= \int g(u,u',u'', \, \dots) ds$,
define $\lb F, G \rb_k = \,< {\tilde \slr}^k \tilde K \nabla F, \nabla G>,\,
k = 0,1,2 , \dots$.  These Poisson brackets are compatible with the (mKdV)
hierarchy in that $all$ of the invariants $I_{2k}$ of the (mKdV) hierarchy
commute with respect to $all$ of the Poisson brackets just listed.  Similarly,
for functionals on
planar  curves one can define the Poisson bracket $\lb F, G \rb_P =
\, < K \nabla F , \nabla G>$.  As a simple corollary of our formula
for the differential of $\slh_P$ in terms of $\slr_P$, one can show that
{\it $\slh_P$ is Poisson with respect to the two brackets
$\lb,\rb_2$ and $\lb,\rb_P$}.

In this paper, we have
reconsidered the equivalence of the planar filament equation and (mKdV);
this equivalence is realized via the elementary geometric transformation
of assigning to a planar curve its curvature function.  In the theory
of integrable systems, there are a number of examples of interesting
pairs of integrable systems which can be shown to be equivalent:
the Kepler problem and geodesic flow on the sphere; the Neumann problem
and geodesic flow on an ellipsoid; the localized induction equation and
the cubic non-linear Schr\"odinger equation (see [Moser],[Has]).  In
each case the transformation relating the two systems is a natural one
from the point of view of differential geometry.  It would be of great
interest to have a unifying theory which could ``explain" the existence
of these several examples.

\ms
\newpage
\centerline{
\epsfbox{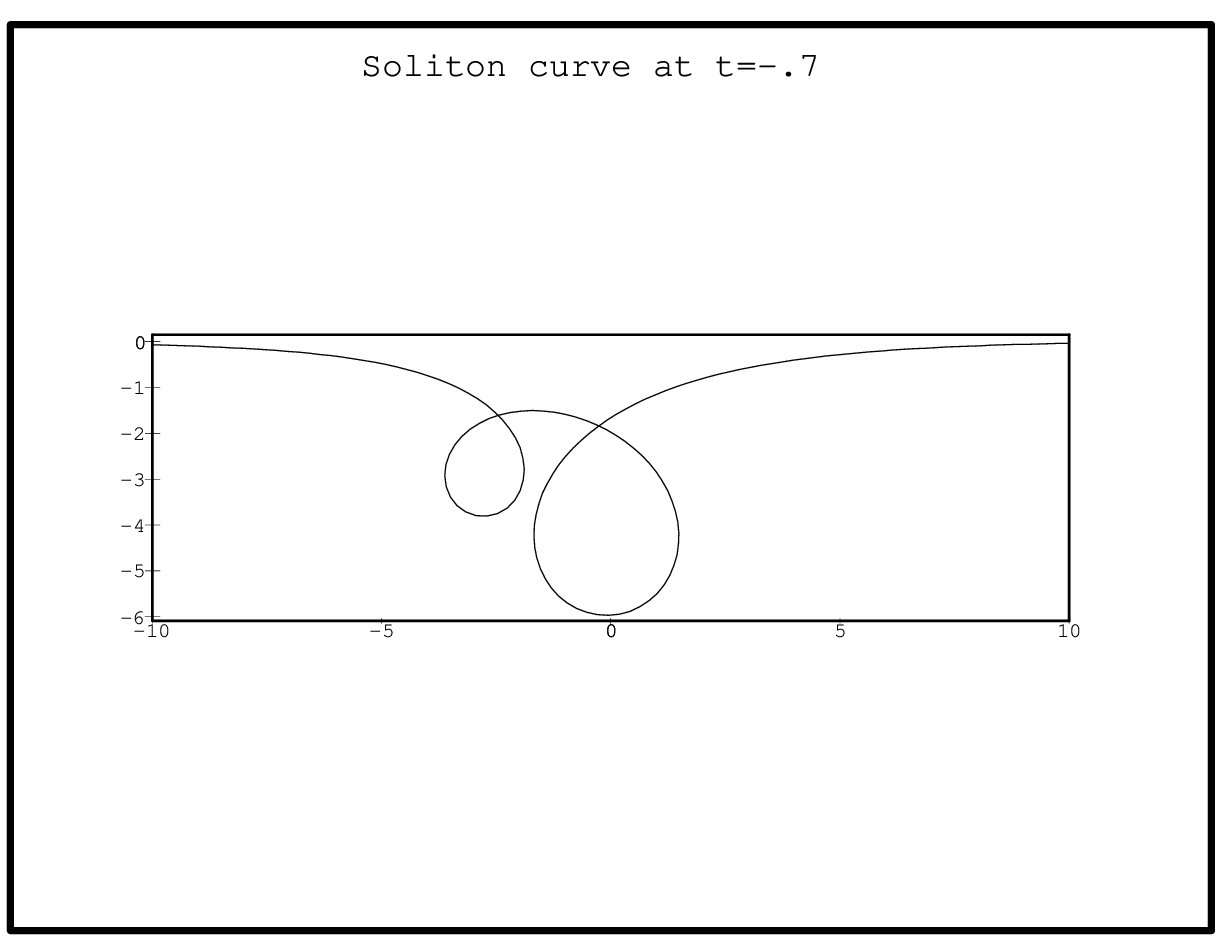}
}
\centerline{
\epsfbox{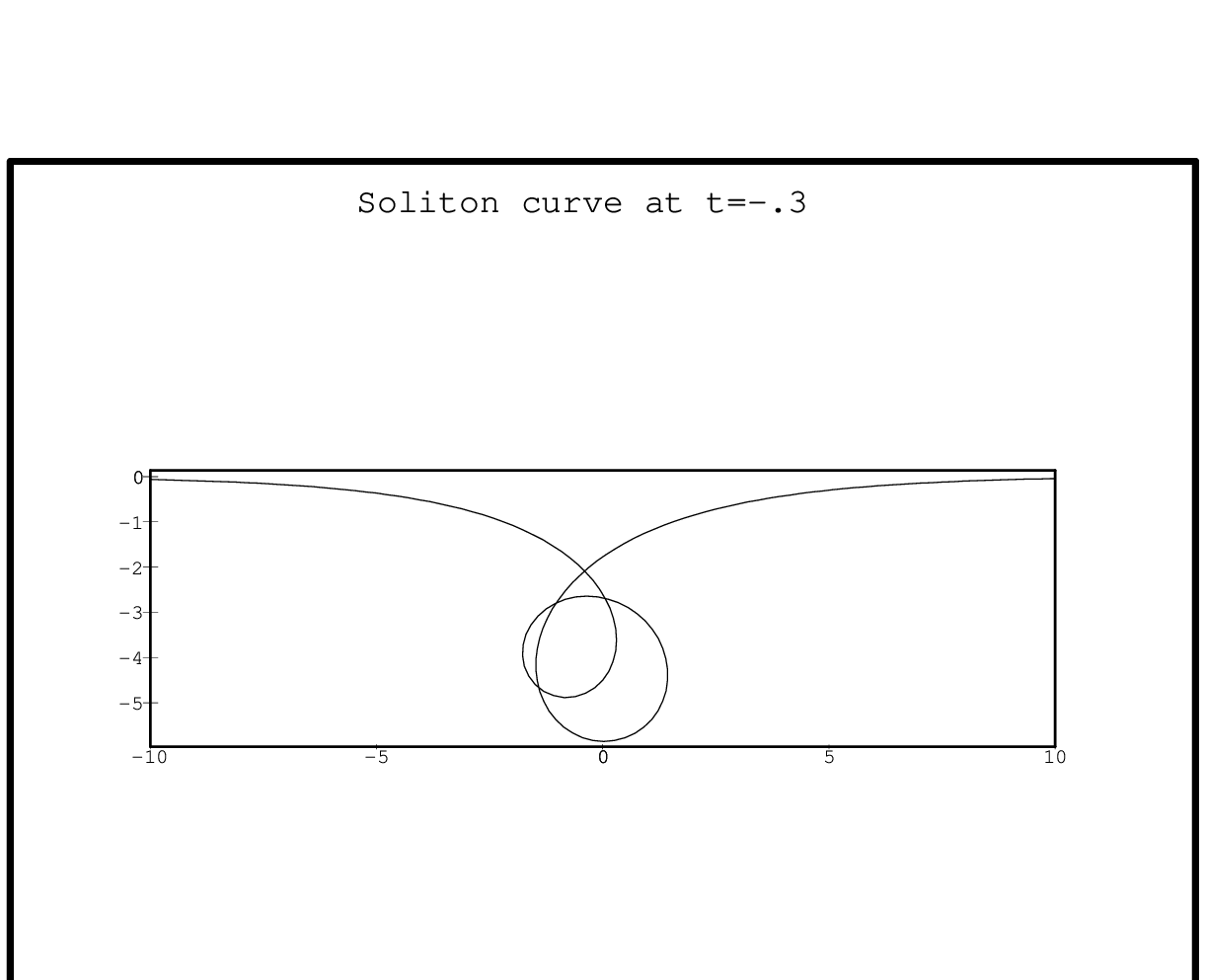}
}
\ms
\newpage
\centerline{
\epsfbox{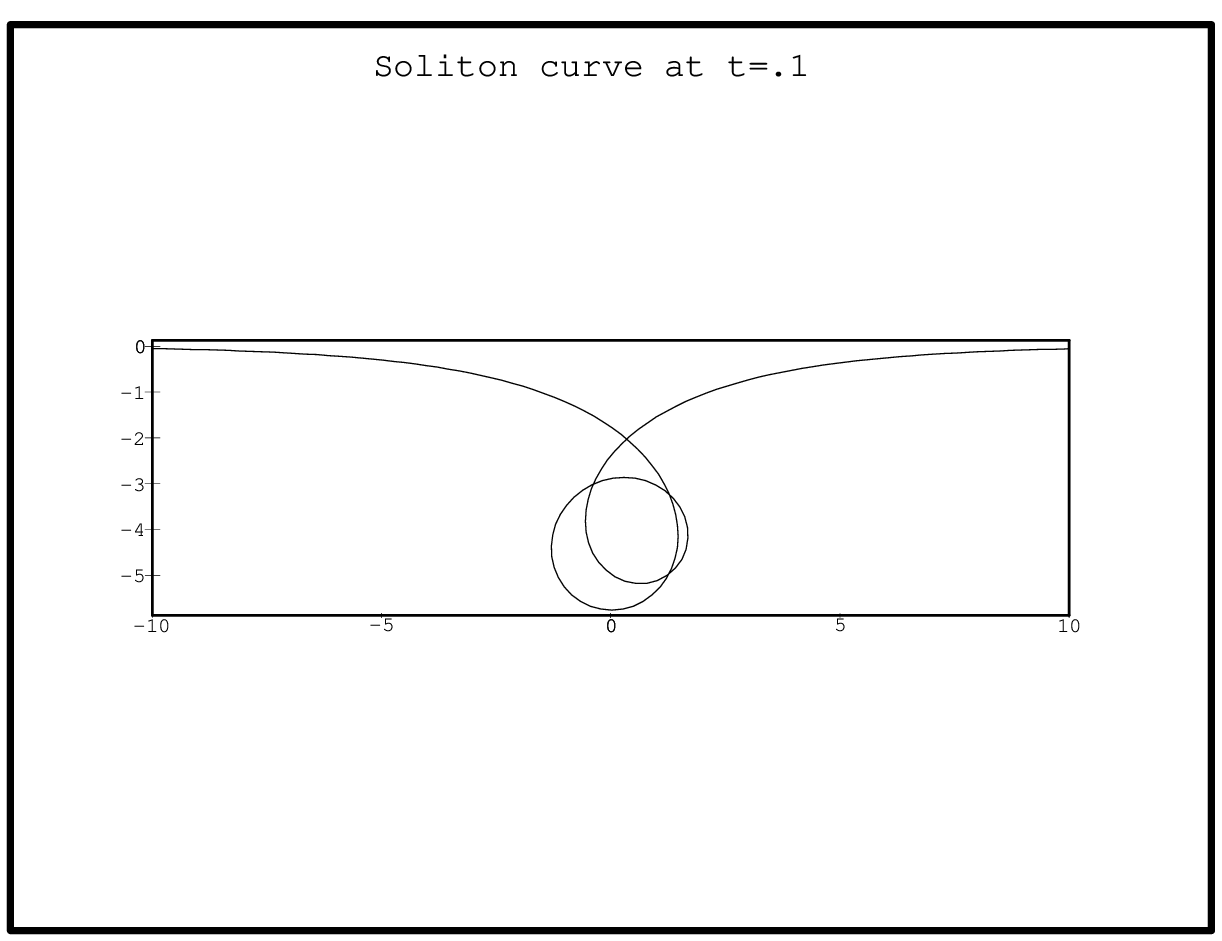}
}
\centerline{
\epsfbox{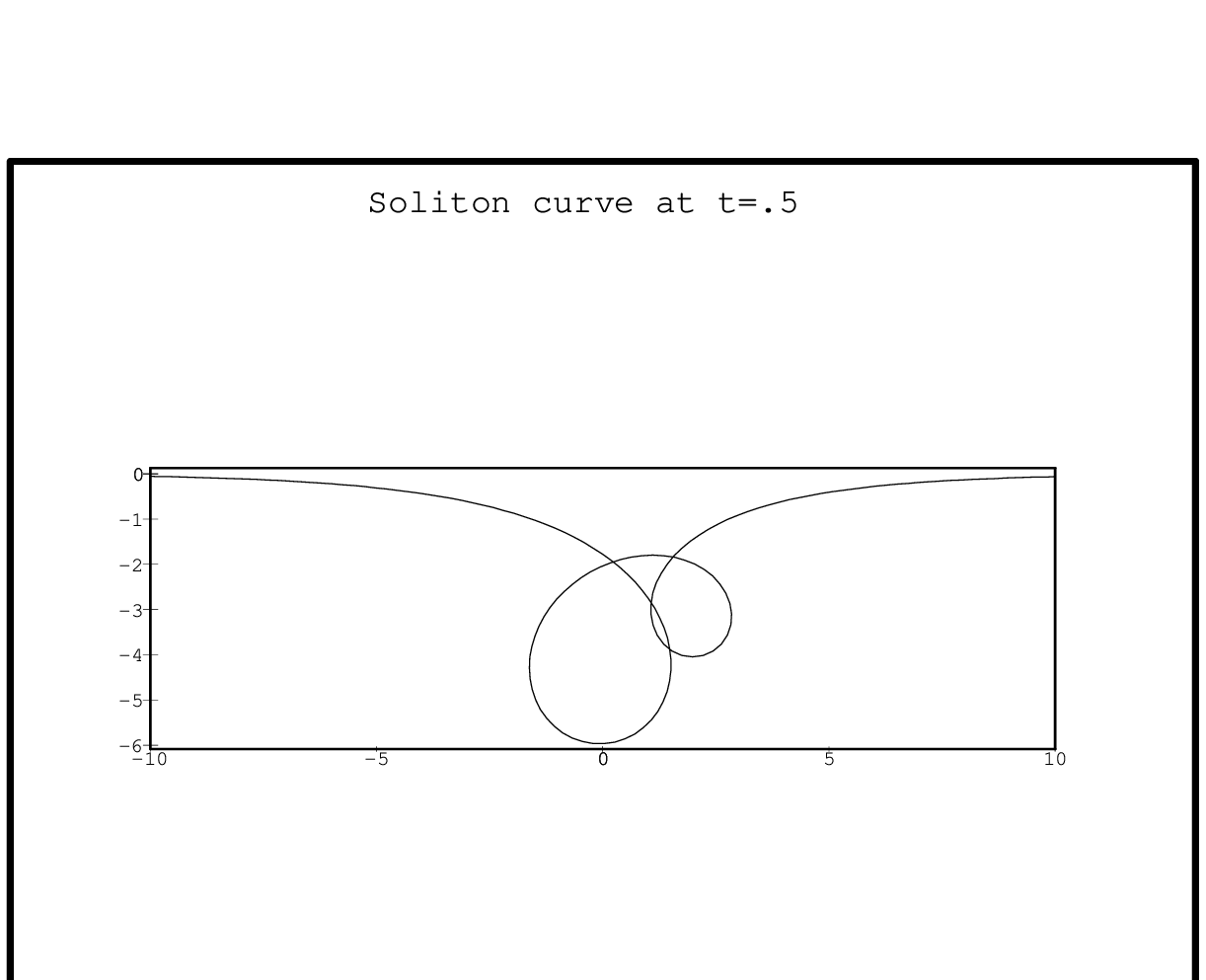}
}
\ms
\newpage

\enddocument